\documentclass[]{spie}  

\usepackage{amsmath,amsfonts,amssymb}
\usepackage{graphicx}
\usepackage[colorlinks=true, allcolors=blue]{hyperref}

\title{Unsupervised brain lesion segmentation from MRI using a convolutional autoencoder}

\author[a]{Hans E. Atlason*}
\author[b,c]{Askell Love}
\author[d]{Sigurdur Sigurdsson}
\author[b,d]{Vilmundur Gudnason}
\author[a,e]{\\Lotta M. Ellingsen}

\affil[a]{Dept. of Electrical and Computer Engineering, University of Iceland, Reykjavik, Iceland}
\affil[b]{Dept. of Medicine, University of Iceland, Reykjavik, Iceland}
\affil[c]{Dept. of Radiology, Landspitali - University Hospital, Reykjavik, Iceland}
\affil[d]{The Icelandic Heart Association, Kopavogur, Iceland}
\affil[e]{Dept. of Electrical and Computer Engineering, The Johns Hopkins University, Baltimore, MD, USA}

\begin{document} 
\maketitle

\begin{abstract}
Lesions that appear hyperintense in both Fluid Attenuated Inversion Recovery (FLAIR) and T2-weighted magnetic resonance images (MRIs) of the human brain are common in the brains of the elderly population and may be caused by ischemia or demyelination. Lesions are biomarkers for various neurodegenerative diseases, making accurate quantification of them important for both disease diagnosis and progression. Automatic lesion detection using supervised learning requires manually annotated images, which can often be impractical to acquire. Unsupervised lesion detection, on the other hand, does not require any manual delineation; however, these methods can be challenging to construct due to the variability in lesion load, placement of lesions, and voxel intensities. Here we present a novel approach to address this problem using a convolutional autoencoder, which learns to segment brain lesions as well as the white matter, gray matter, and cerebrospinal fluid by reconstructing FLAIR images as conical combinations of softmax layer outputs generated from the corresponding T1, T2, and FLAIR images. Some of the advantages of this model are that it accurately learns to segment lesions regardless of lesion load, and it can be used to quickly and robustly segment new images that were not in the training set. Comparisons with state-of-the-art segmentation methods evaluated on ground truth manual labels indicate that the proposed method works well for generating accurate lesion segmentations without the need for manual annotations.
\end{abstract}


\section{INTRODUCTION}

Lesions that correspond to white matter (WM) hyperintensities in Fluid Attenuated Inversion Recovery (FLAIR) and T2-weighted (T2-w) magnetic resonance images (MRIs) have been shown to be positively associated with age and neurodegenerative diseases \cite{black2009understanding}. Such lesions are observed in 96\% of volunteers over the age of 65 \cite{black2009understanding}. An accurate segmentation of these lesions is important in large-scale imaging studies of the elderly, both because the lesions are important biomarkers to characterize brain diseases and because they can interfere with the segmentation of other brain structures if not accounted for in automated segmentation approaches. 
The currently accepted gold standard in lesion segmentation is manual delineation by an expert in neuroanatomy. However, acquiring such delineations is both time consuming and expensive and studies have shown that human raters can have great intra- and inter-rater variability \cite{carass2017longitudinal}.

Multiple automatic lesion segmentation methods have been developed that can be categorized into supervised and unsupervised methods. Supervised methods rely on training sets that consist of MRIs and corresponding manually delineated lesions. Although supervised methods with convolutional neural networks (CNNs) have shown to achieve state-of-the-art performance on various data sets \cite{carass2017longitudinal, li2018fully,  roy2018multiple}, manual delineations often don't exist when analyzing new data sets and creating them may not be practical. Unsupervised methods typically involve modeling of MRI brain tissue intensities. This can be challenging because tissue intensities of MRIs are not always consistent within an image (e.g., due to inhomogeneity artifacts) or between different images (e.g., due to scanner differences). Artifacts or poor skull-stripping can lead to high-intensity regions in FLAIR images that could potentially be incorrectly classified as lesions. Furthermore, the number of lesions and their location is unknown beforehand and can vary greatly between subjects, with some subjects having no lesions while others have an abundance of them. When using such automated approaches to process large data sets of brain MRIs, one must be able to trust that lesion boundaries are consistent within and across subjects, and that the methods are reliable and robust to different sizes and numbers of lesions.

Various methods have been proposed for unsupervised lesion segmentation, e.g. methods that use tissue segmentation to obtain lesions\cite{garcia2011trimmed, schmidt2012automated} (usually as outliers of tissue \cite{llado2012segmentation}), and approaches that use only specific lesion properties\cite{tomas2015model,llado2012segmentation}. Furthermore, lesions can be detected as outliers of “pseudo-healthy” synthesized images\cite{bowles2017brain, baur2018deep}, however, a necessary requirement for these approaches to work is to have a training data set with healthy brains to model normality, such that lesions can be detected either as outliers or as results of large reconstruction errors \cite{bowles2017brain, baur2018deep, pawlowski2018unsupervised}. This is not the case when analyzing data sets of elderly subjects, where the variability in lesion load (both number and size of lesions) can be extremely high. The proposed method addresses this problem, by enabling unsupervised tissue and lesion classification with a convolutional autoencoder without the need to model lesions as anomalies in the brain. We propose to model the intensities of FLAIR images as a conical combination of the segmentation of WM, gray matter (GM), cerebrospinal fluid (CSF), and lesions, estimated by a convolutional autoencoder from the corresponding T1, T2 and FLAIR images. Some of the advantages of this model are that it accurately learns to segment lesions regardless of lesion load, and it can be used to quickly and robustly segment new images that were not in the training set. We will hereafter refer to the proposed method as the Segmentation Auto-Encoder (SegAE).

\section{METHOD}
\subsection{Data and preprocessing}
The AGES-Reykjavik Study \cite{sigurdsson2012brain} cohort comprises 5764 participants (female and male, age 66-93), 4811 of which underwent brain MRI.  The MRIs were acquired using a dedicated General Electrics 1.5-Tesla Signa Twinspeed EXCITE system with a multi-channel phased array head cap coil. T1-weighted (T1-w) images (TE = 8 ms, TR = 21 ms) with 0.94$\times$0.94$\times$1.5 mm\textsuperscript{3} voxel size; T2-w fast spin echo sequence (TE = 90 ms, TR = 3220 ms); and FLAIR sequence (TE = 100ms, TR = 8000ms) with 0.86$\times$0.86$\times$3.0 mm\textsuperscript{3} voxel size.

For developmental purposes, we randomly selected 50 subjects from the AGES-Reykjavik study; 30 subjects for training, 5 for validation of model parameters, and 15 for testing. The WM lesions in the test images were manually annotated by an experienced neuroradiologist. The images used for validation were used to determine model architecture and hyperparameters based on visual inspection.

The images were preprocessed using standard preprocessing procedures: Resampling to 0.8$\times$0.8$\times$0.8 mm\textsuperscript{3} voxel size, correction for signal non-uniformity~\cite{tustison2010n4itk} in the T1-w and T2-w images (see subsection \ref{sec:postprocessing} for why FLAIR was omitted), rigid registration to the MNI-ICBM152 template~\cite{fonov2009unbiased}, and skull removal~\cite{Roy_NI17}.

\subsection{Network architecture and training procedure}
The proposed method, SegAE, makes use of an autoencoder with fully convolutional layers on three resolution scales. The input to SegAE consists of three-dimensional (3D) patches of size 80x80x80 voxels each, from T1-w, T2-w, and FLAIR images. The autoencoder is then trained to minimize the loss L:
$$
L=(Y^p-\hat{Y}^p )^2,
$$
where $Y$ is a FLAIR patch, $\hat{Y}$ is the corresponding reconstructed FLAIR patch from the autoencoder, and $p=3$ to give more weight to the hyperintense lesions. A brainmask~\cite{Roy_NI17} is applied before the loss so the background is not considered. The last layer is a one-by-one convolution constrained to have positive weights and no bias\cite{palsson2018hyperspectral}. The input to the last layer has a sum-to-one constraint enforced with a softmax activation function. This is an ill-posed problem, so regularization is needed. We want to push the sum-to-one layer to generate binary segmentations. This is done by constraining the input to the softmax function to be in a certain range (0-200) along the channel axis.

Patches were normalized to have unit mean WM before training, using a WM segmentation mask. We used FreeSurfer\cite{fischl2012freesurfer} to generate this mask, however, any tissue segmentation method can be used. To make the network less sensitive to image normalization the training patches of each channel were multiplied by a constant drawn from a Gaussian distribution with a mean vale of 1.0 and standard deviation of 0.5 during training. Furthermore,  Gaussian noise with zero mean and a standard deviation of 0.05 was added to the input during training for noise robustness. The network architecture for the convolutional autoencoder can be seen in Figure~\ref{f:cnn_model}. Patches from the 30 training images were acquired with a stride of 40, and a GTX1080 Ti GPU was used to train the network for 50 epochs with a learning rate of 0.001 using the Adam optimizer~\cite{kingma2014adam} with $\beta_1=0.99$ and $\beta_2=0.99999$ and a batch size of one. Leaky rectified linear unit (LReLU) activation functions had a slope of 0.1 for the negative part. Hyperparameters where chosen by trial and error.

\subsection{Prediction and post-processing}
\label{sec:postprocessing}
\begin{figure}[!tb]
\centering
\includegraphics[width=0.8\textwidth]{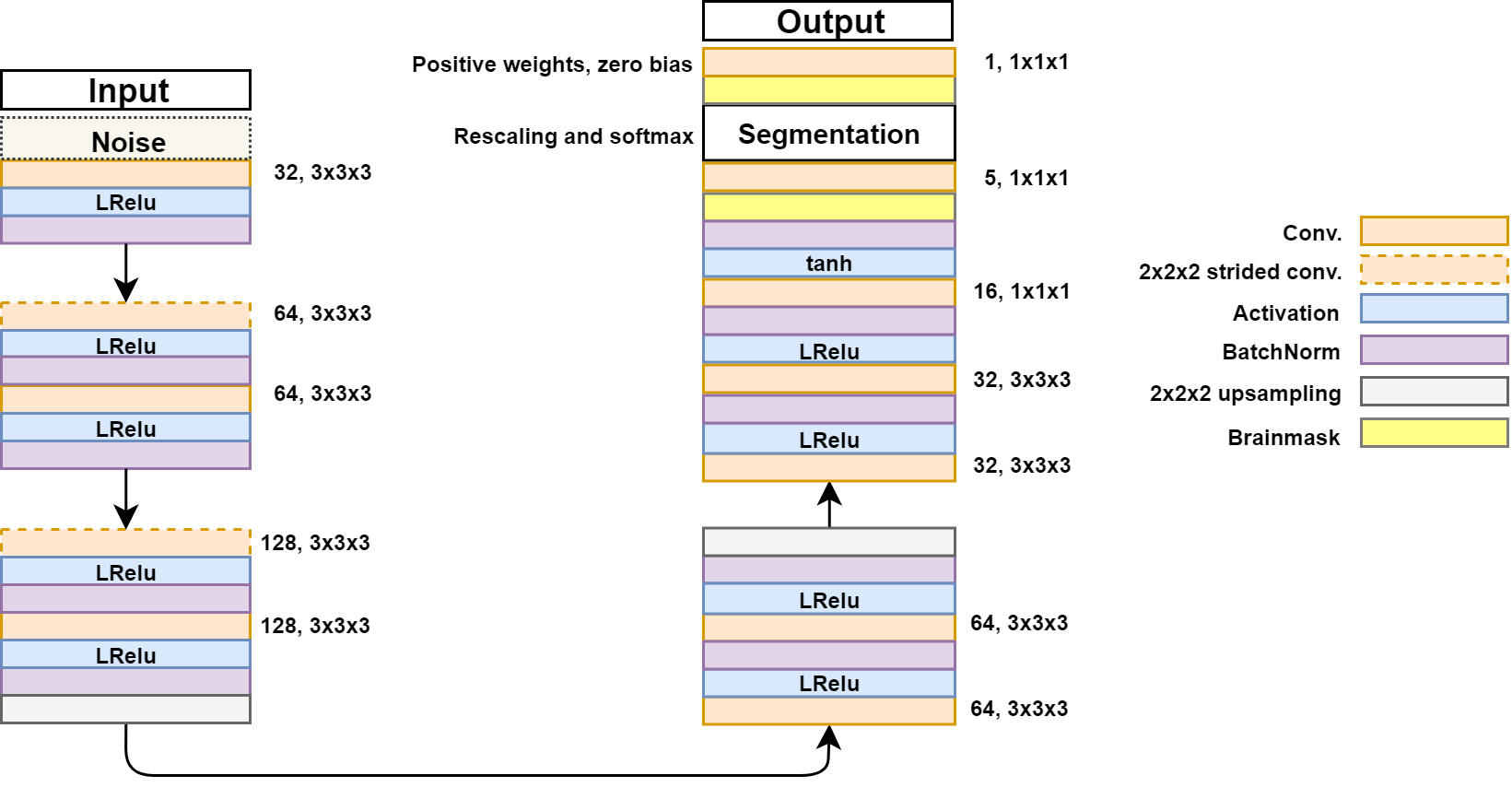}
\caption{The proposed convolutional autoencoder architecture. The input is 3D patches from T1, T2, and FLAIR MRIs. The number of filters and kernel size are shown next to the corresponding convolutional layers. The Segmentation layer denotes a rescaling of the activations and an application of a softmax function. The final convolutional layer is restricted to have positive weights and zero bias for the output reconstruction of the FLAIR patch to be a conical combination of the Segmentation layer.}
\label{f:cnn_model}
\end{figure}

After training, one of the five Segmentation output volumes (see Figure~\ref{f:cnn_model}) that corresponds to lesion segmentation was used for prediction, and other outputs were discarded. Prediction was performed with a stride of 20, and patches were assembled using the average of overlapping voxels. A threshold of 0.5 was used to binarize the average predictions.
Initially, N4 \cite{tustison2010n4itk} was used for inhomogeneity correction in the T1, T2 and FLAIR images. However, we observed that FLAIR lesions tended to degrade after N4 correction. Therefore, we did not use N4 correction on the FLAIR images used for training and testing SegAE, resulting in some high intensity inhomogeneity artifacts appearing mainly in the anterior and posterior parts of the cortical gray matter.

Following the autoencoder, an additional post-processing step was included involving multiplying the result from SegAE with two morphologically eroded brainmasks. 
The first brainmask was obtained by morphologically eroding the skullstripping mask with a 3x3x3 cube in 10 iterations. The second brainmask was obtained from FreeSurfer (binarized segmentation excluding sulcal CSF) and morphologically eroded with a 3x3x3 cube in 2 iterations. The post-processing removed most of the inhomogeneity artifacts from the validation subjects, as well as artifacts due to failures in skullstripping. Although we used FreeSurfer to generate the brainmask that excludes sulcal CSF, any algorithm capable of producing such a brainmask would work.


\section{RESULTS AND DISCUSSION}
\begin{figure}[!tb]
\centering
\begin{tabular}{ccccccc}
\textbf{FLAIR image} & \textbf{Supervised} & \textbf{FreeSurfer} & \textbf{S3DL} &
\textbf{SegAE} &  \textbf{Manual}\\
\includegraphics[width=0.14\textwidth, clip=True, trim=5cm 5cm
5cm 5cm, keepaspectratio]{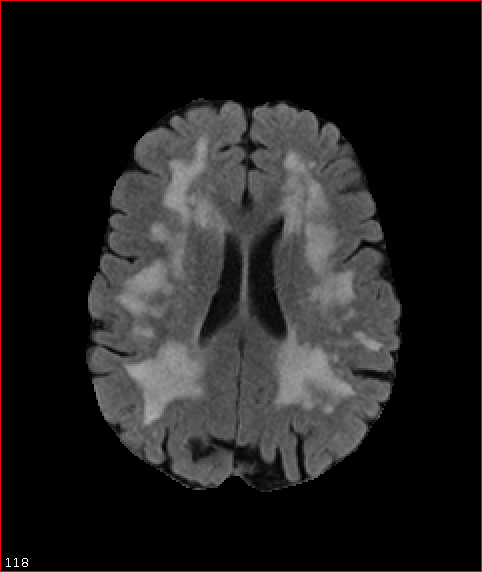} &
\includegraphics[width=0.14\textwidth, clip=True, trim=5cm 5cm
5cm 5cm]{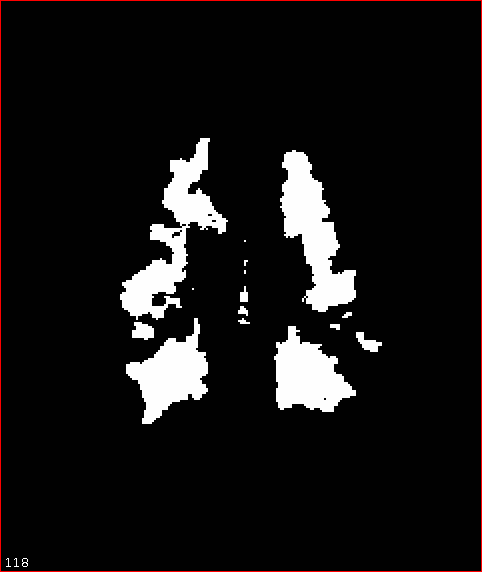} &
\includegraphics[width=0.14\textwidth, clip=True, trim=5cm 5cm
5cm 5cm]{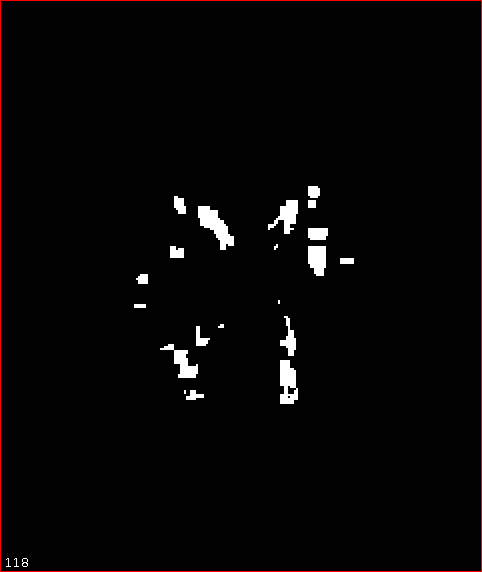} &
\includegraphics[width=0.14\textwidth, clip=True, trim=5cm 5cm
5cm 5cm]{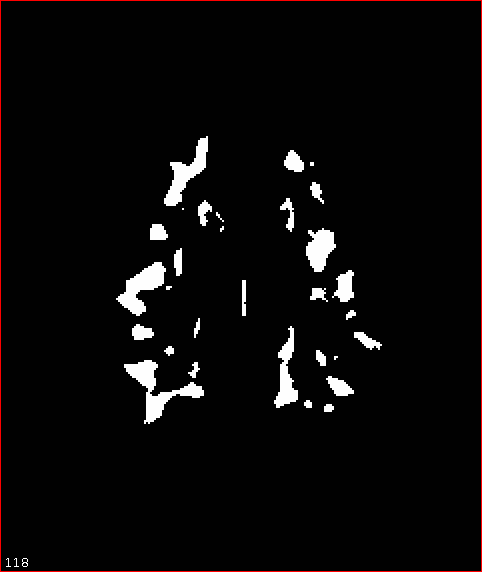} &
\includegraphics[width=0.14\textwidth, clip=True, trim=5cm 5cm
5cm 5cm]{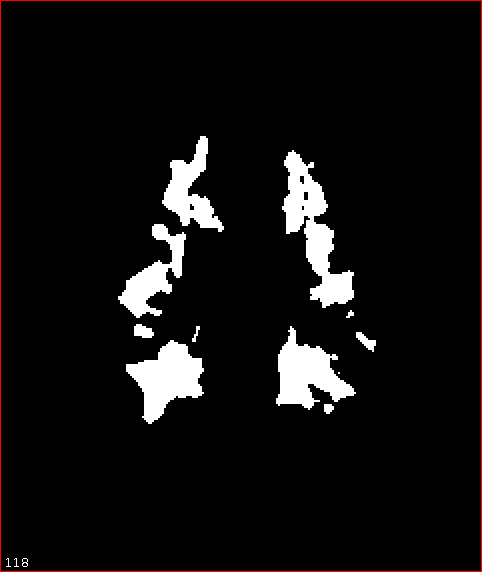} &
\includegraphics[width=0.14\textwidth, clip=True, trim=5cm 5cm
5cm 5cm]{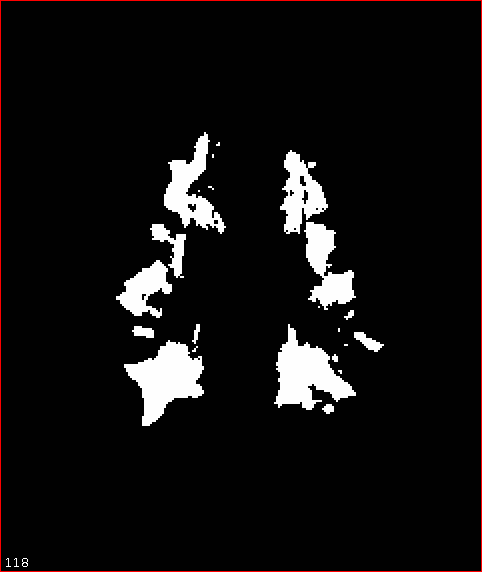} 
\\
\includegraphics[width=0.14\textwidth, clip=True, trim=5cm 5cm
5cm 5cm]{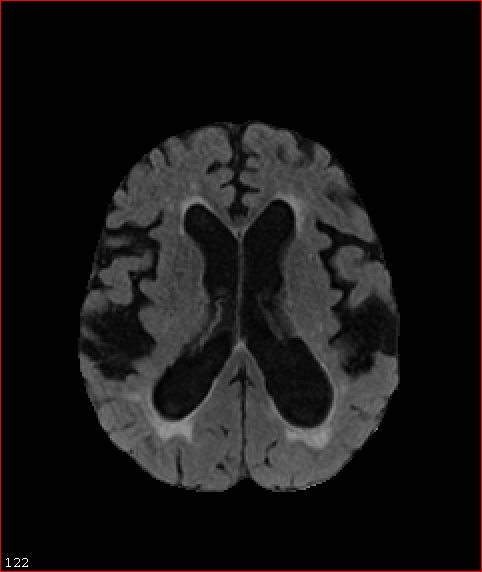} &
\includegraphics[width=0.14\textwidth, clip=True, trim=5cm 5cm
5cm 5cm]{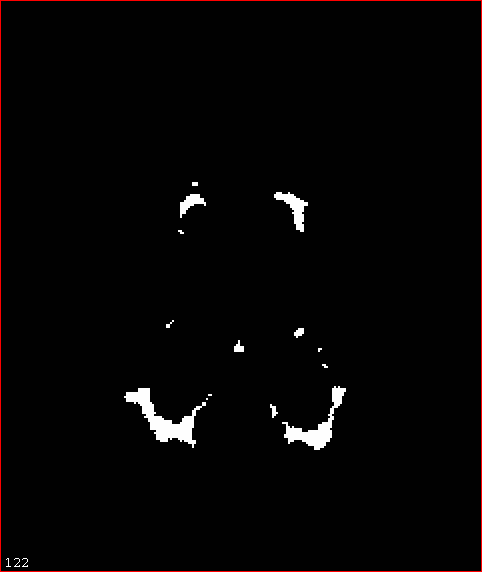} &
\includegraphics[width=0.14\textwidth, clip=True, trim=5cm 5cm
5cm 5cm]{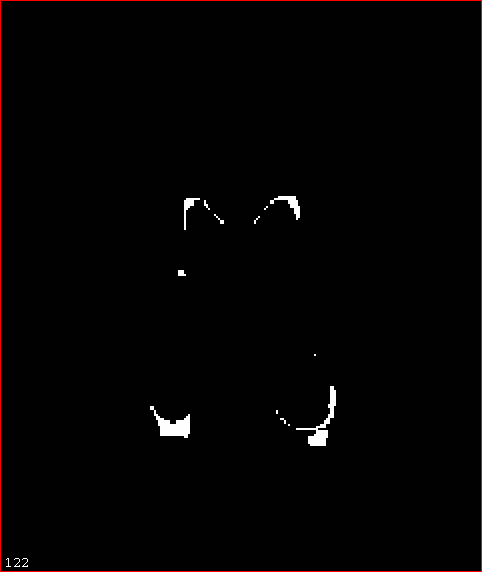} &
\includegraphics[width=0.14\textwidth, clip=True, trim=5cm 5cm
5cm 5cm]{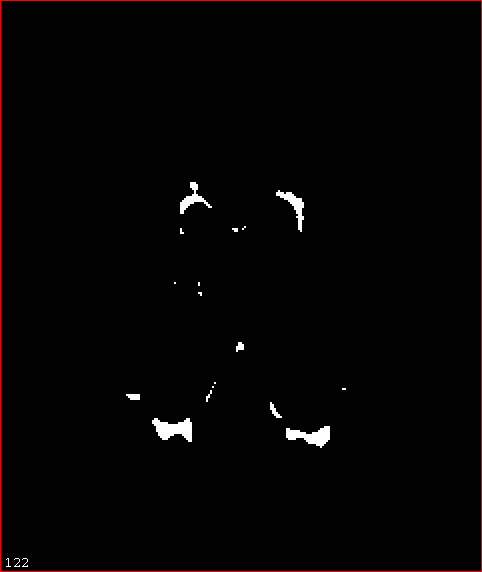} &

\includegraphics[width=0.14\textwidth, clip=True, trim=5cm 5cm
5cm 5cm]{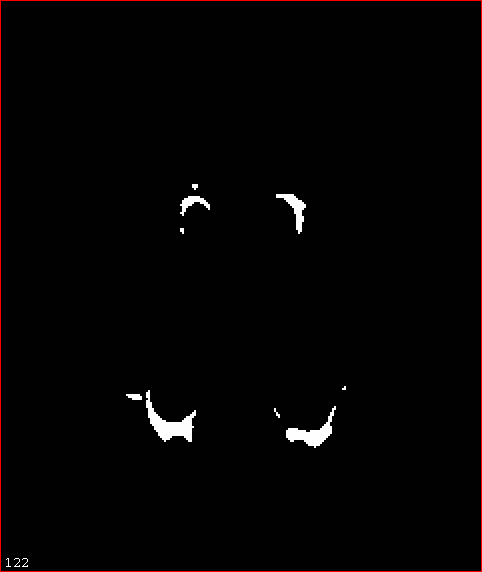} &
\includegraphics[width=0.14\textwidth, clip=True, trim=5cm 5cm
5cm 5cm]{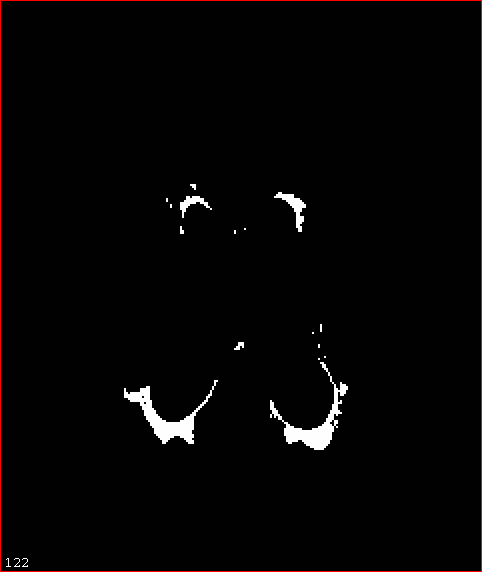}
\\
%
%
\end{tabular}
\caption{Visual comparison of the proposed method and three state-of-the-art methods with a manual rater for two different subjects, one with a high lesion load (top row) and one with a smaller lesion load but with enlarged ventricles (bottom row).}

\label{f:lesions}
\end{figure}

The white matter lesions in a total of 15 subjects were manually delineated by a neuroradiologist to be used as ground truth lesion segmentations for evaluation of the proposed method. We compared the proposed method with three state-of-the-art segmentation methods: 1) A supervised tissue segmentation developed for the AGES-Reykjavik data set, created with an artificial neural network classifier in the four dimensional intensity space defined by FLAIR, T1-w, T2-w and Proton Density weighted images and trained on 11 manually annotated subjects\cite{sigurdsson2012brain}; 2) the patch-based Subject Specific Sparse Dictionary Learning (S3DL) method \cite{roy2015subject}, which takes FLAIR and T1-weighted images as input for lesion segmentation as well as three manually annotated atlases; and 3) the whole brain segmentation method FreeSurfer \cite{fischl2012freesurfer}, which only takes a T1-weighted image as input, but is included in the comparison due to its widespread use. For each of the methods above,  the preprocessing steps were as described in their associated publications. A visual comparison of the methods is shown for two subjects in Figure~\ref{f:lesions}. The top row in Figure~\ref{f:lesions} demonstrates that SegAE can accurately segment lesion boundaries of the test image with the largest lesion load. The bottom row shows an example of a subject with a smaller lesion load but with enlarged ventricles. 

\begin{figure}[!tb]
\centering
\setlength{\tabcolsep}{0.1em}
\begin{tabular}{cccc}
\textbf{AVD} & \textbf{Dice} & \textbf{PPV} &
\textbf{TPR}\\
\includegraphics[width=0.245\textwidth, clip=True, trim=0.5cm 0cm
1.4cm 1cm]{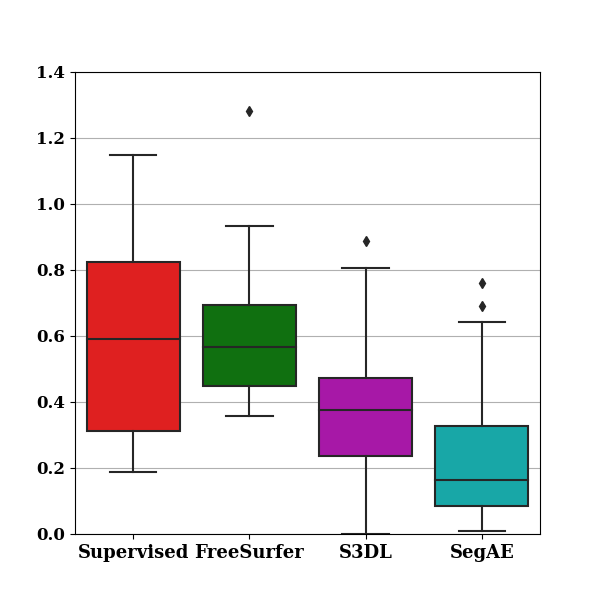} &
\includegraphics[width=0.245\textwidth, clip=True, trim=0.5cm 0cm
1.4cm 1cm]{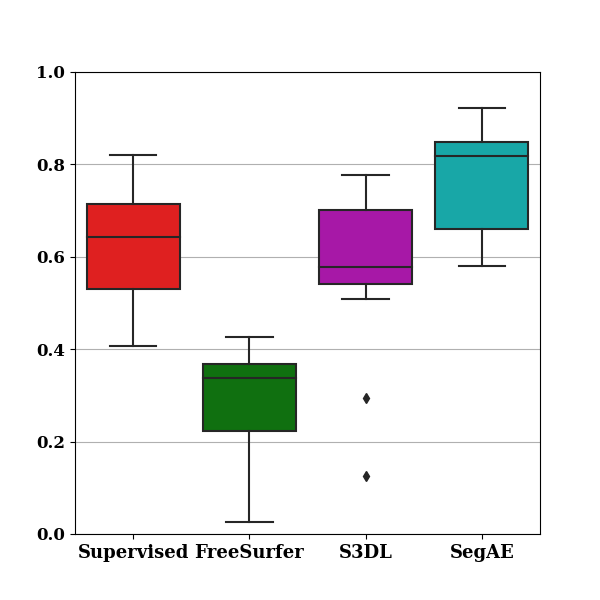} &
\includegraphics[width=0.245\textwidth, clip=True, trim=0.5cm 0cm
1.4cm 1cm]{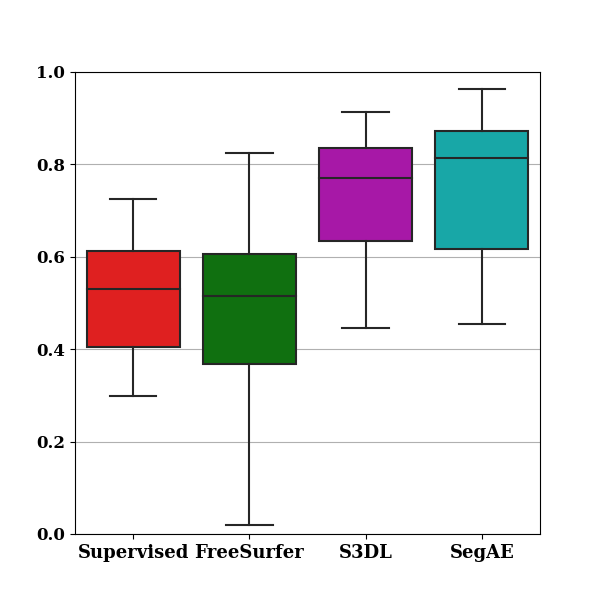} &
\includegraphics[width=0.245\textwidth, clip=True, trim=0.5cm 0cm
1.4cm 1cm]{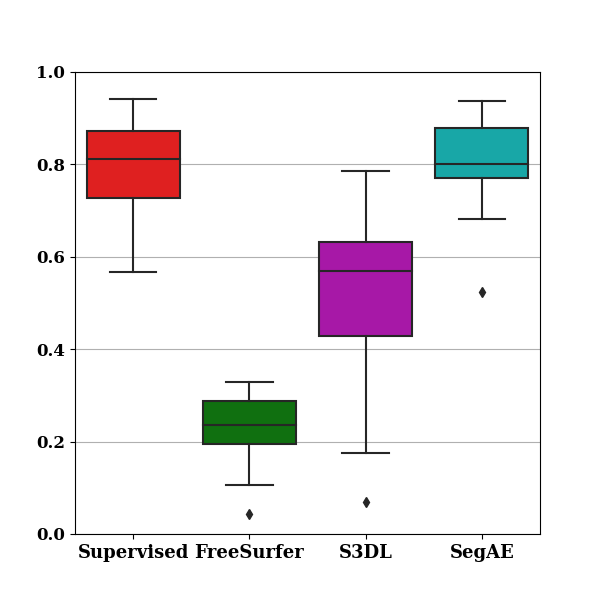}
\\
\end{tabular}

\caption{Boxplots showing the AVD, Dice, PPV, and TPR scores of Supervised (red), FreeSurfer (green), S3DL (magenta), and SegAE (cyan), respectively.}
\label{f:boxplot}
\end{figure}

\begin{figure}[!tb]
\centering
\setlength{\tabcolsep}{0.1em}
\begin{tabular}{cccc}
\textbf{Supervised} & \textbf{FreeSurfer} & \textbf{S3DL} &
\textbf{SegAE}\\
\includegraphics[width=0.245\textwidth, clip=True, trim=0cm 0cm
1.4cm 1cm]{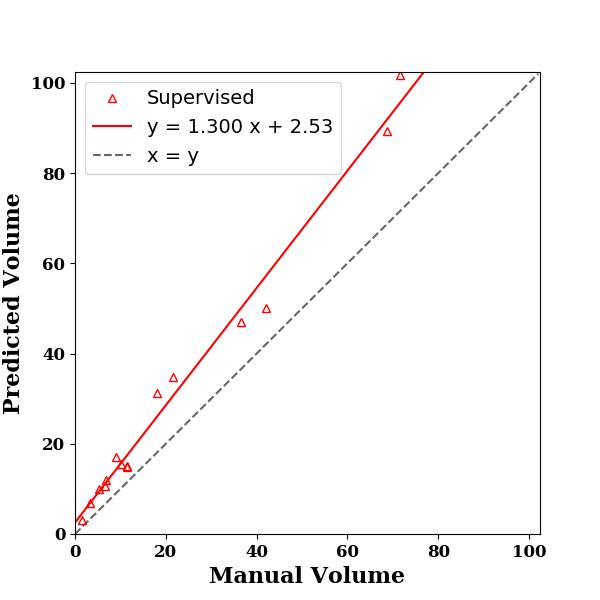} &
\includegraphics[width=0.245\textwidth, clip=True, trim=0cm 0cm
1.4cm 1cm]{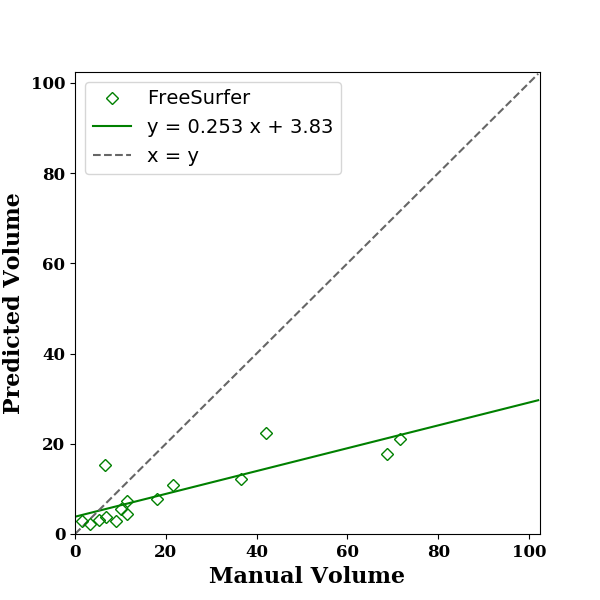} &
\includegraphics[width=0.245\textwidth, clip=True, trim=0cm 0cm
1.4cm 1cm]{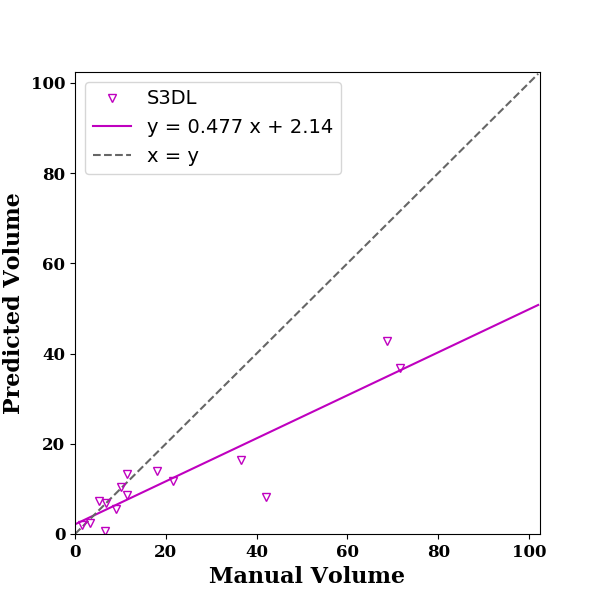} &
\includegraphics[width=0.245\textwidth, clip=True, trim=0cm 0cm
1.4cm 1cm]{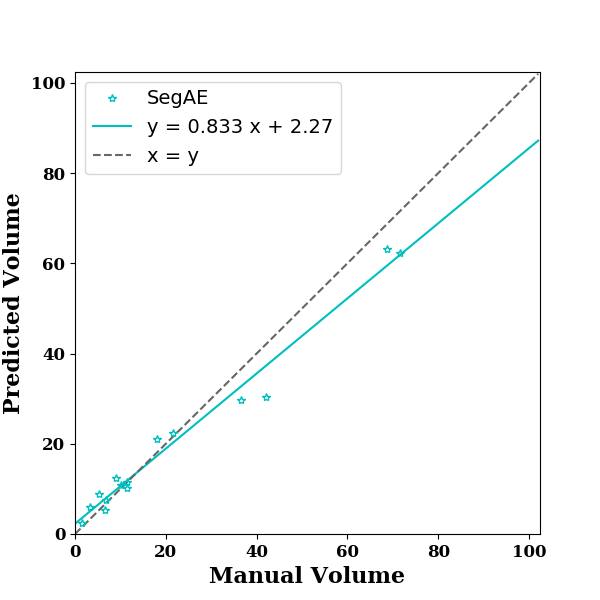}
\\
\end{tabular}

\caption{Predicted lesion volumes versus manual lesion volumes for the 4 methods. The solid lines show a linear fit of the points and the dashed black line has unit slope. Numbers are in cm$^3$.}
\label{f:scatter_volumes}
\end{figure}

The Dice score, Positive Predictive Value (PPV), True Positive Rate (TPR), and Absolute Volume Difference (AVD) were used as quantitative evaluation metrics, as described in Carass et al. \cite{carass2017longitudinal}. 
Figure~\ref{f:boxplot} shows box plots of AVD, Dice, PPV, and TPR scores for each method.
The mean values and standard deviations are reported in Table \ref{t:dices}.
SegAE achieves the lowest average AVD of 0.255, the highest Dice score of 0.766 (p $<$ 0.005), the highest average PPV of 0.757, and the highest average TPR of 0.802. The p-values from a paired Wilcoxon signed-rank test comparing the AVD, Dice, PPV, and TPR scores of SegAE to the Supervised method, FreeSurfer, and S3DL can be seen in Table~\ref{t:pvals}.
We compared the predicted lesion volumes to the manual lesion volumes in Figure~\ref{f:scatter_volumes}. The solid lines show a linear fit of the points and the dashed black lines have a unit slope. We observe that the Supervised method systematically overestimates lesion volumes of both small and large lesions. FreeSurfer and S3DL seem to underestimate most lesion volumes in this data set. SegAE does not show such a bias, although the predicted volumes of the four largest lesions are slightly smaller than the manual volumes. 
Table~\ref{t:slopes} shows the slopes and intercepts of the linear fits between the five methods and the manual volumes.

\begin{table}[!tb]
\caption{The mean and standard deviation of Absolute Volume Difference (AVD), Dice, Positive Predictive Value (PPV), and True Positive Rate (TPR) for the proposed method SegAE, and three alternative methods. Asterisk (*) denotes that a value is significantly different (p $<$ 0.005) from SegAE, and bold figures denote the best result.}
\begin{center}
\begin{tabular}{lcccc } 
 & AVD & Dice & PPV & TPR \\
 \hline
Supervised & 0.605 ($\pm$ 0.298)* & 0.622 ($\pm$ 0.127)* & 0.517 ($\pm$ 0.134)* & 0.794 ($\pm$ 0.107)\\
FreeSurfer & 0.611 ($\pm$ 0.238)* & 0.294 ($\pm$ 0.113)* & 0.479 ($\pm$ 0.224)* & 0.226 ($\pm$ 0.081)* \\
S3DL & 0.370 ($\pm$ 0.241) & 0.571 ($\pm$ 0.167)* & 0.735 ($\pm$ 0.144) & 0.516 ($\pm$ 0.193)*\\
SegAE & \textbf{0.255 ($\pm$ 0.240)} & \textbf{0.766 ($\pm$ 0.114)} & \textbf{0.757 ($\pm$ 0.169)} & \textbf{0.802 ($\pm$ 0.099)}\\
\hline
\end{tabular}
\end{center}
\label{t:dices}
\end{table}

\begin{table}[!tb]
\caption{The p-values of a paired Wilcoxon signed-rank test (without correction for multiple comparisons) comparing the AVD, Dice, PPV, and TPR scores of SegAE and the Supervised method, FreeSurfer, and S3DL.}
\begin{center}
\begin{tabular}{lcccc } 
 & AVD & Dice & PPV & TPR \\
 \hline
SegAE vs. Supervised & 0.0008  & 0.0007  & 0.0007  & 0.6909 \\
SegAE vs. FreeSurfer & 0.0064  & 0.0007  & 0.0007  & 0.0007  \\
SegAE vs. S3DL   & 0.1728  & 0.0007  & 0.0691 & 0.0007\\
\hline
\end{tabular}
\end{center}
\label{t:pvals}
\end{table}

\begin{table}[!tb]
\caption{Slopes and intercepts from Figure~\ref{f:scatter_volumes}.}
\begin{center}
\begin{tabular}{lcccc} 
 & Supervised & FreeSurfer & S3DL & SegAE \\
 \hline
Slope & 1.300  & 0.253 & 0.477 & 0.833 \\
Intercept & 2.533  &  3.828 & 2.144 & 2.273\\

\hline
\end{tabular}
\end{center}
\label{t:slopes}
\end{table}

\section{CONCLUSIONS}
We have presented SegAE, a convolutional autoencoder architecture that can be used for unsupervised segmentation of WM brain lesions from MRIs. We note that an important byproduct of the approach is a classification of the brain into WM, GM, and CSF (results not reported). The model was trained and evaluated on a data set with a high variability in lesion load. The lesion segmentation was evaluated on 15 manually labeled subjects and compared with three state-of-the-art tissue and lesion segmentation methods. SegAE achieves the lowest mean AVD, the highest Dice (significantly higher than all three methods, p $<$ 0.005), the highest mean PPV, and the highest mean TPR on these test images. Table~\ref{t:pvals} shows the p-values comparing the three methods to SegAE.

Future work includes using a scale-invariant loss function to stabilize training, incorporating the other imaging modalities into the loss, exploring better methods of pre- and post-processing to prevent over-segmentation due to image artifacts, and a more extensive evaluation of the proposed method on a larger data set of manually delineated subjects.


\section{Acknowledgements}
This work was supported by RANNIS (The Icelandic Centre for Research) through grant 173942-051.

\bibliography{papers} 
\bibliographystyle{spiebib} 

\end{document}